\documentclass[prl,showpacs,amsmath,amssymb,floatfix,twocolumn]{revtex4}

\usepackage{graphicx}% Include figure files
\usepackage{dcolumn}% Align table columns on decimal point
\usepackage{bm}% bold math
\usepackage{psfrag}
\usepackage{subfigure}

\newcommand{\be}{\begin{eqnarray}}
\newcommand{\ee}{\end{eqnarray}}
\newcommand{\bn}{\begin{eqnarray*}}
\newcommand{\en}{\end{eqnarray*}}

\newcommand{\nn}{\nonumber \\}

\renewcommand{\vec}[1]{\mbox{\boldmath$#1$}}
\renewcommand{\d}{\mbox{\rm d}}

\renewcommand{\th}{\ensuremath{\theta}}

\newcommand{\vph}{\ensuremath{\varphi}}
\newcommand{\al}{\ensuremath{\alpha}}
\newcommand{\bt}{\ensuremath{\beta}}
\newcommand{\sg}{\ensuremath{\sigma}}
\newcommand{\gm}{\ensuremath{\gamma}}

\newcommand{\lm}{\ensuremath{\lambda}}

\newcommand{\Sg}{\ensuremath{\Sigma}}
\newcommand{\Gm}{\ensuremath{\Gamma}}
\newcommand{\Om}{\ensuremath{\Omega}}

\newcommand{\Cx}{\ensuremath{\hat{x}}}
\newcommand{\Cy}{\ensuremath{\hat{y}}}
\newcommand{\Cz}{\ensuremath{\hat{z}}}

\newcommand{\Cj}{\ensuremath{\hat{\jmath}}}

\newcommand{\ze}{\ensuremath{\hat{0}}}

\newcommand{\pvec}{\ensuremath{\vec{p}}}

\newcommand{\kvec}{\ensuremath{\vec{k}}}

\newcommand{\Lvec}{\ensuremath{\vec{L}}}

\newcommand{\rvec}{\ensuremath{\vec{r}}}

\newcommand{\alvec}{\ensuremath{\vec{\al}}}
\newcommand{\sgvec}{\ensuremath{\vec{\sg}}}
\newcommand{\Sgvec}{\ensuremath{\vec{\Sg}}}

\newcommand{\lt}{\ensuremath{\left}}
\newcommand{\rt}{\ensuremath{\right}}

\newcommand{\nabvec}{\ensuremath{\vec{\nabla}}}
\newcommand{\PhiG}{\ensuremath{\Phi_{\rm G}}}

\renewcommand{\k}{\ensuremath{k_0}}

\renewcommand{\d}{\mbox{\rm d}}

\begin{document}

%\preprint{APS/123-QED}

%\title{Manuscript Title:\\with Forced Linebreak}% Force line breaks with \\

%\author{Ann  Author}
%\affiliation{%

%\affiliation{


%\preprint{APS/123-QED}

\title{Can gravity distinguish between Dirac and Majorana neutrinos?}%

%\author{Dinesh Singh}
%\email{singhd@uregina.ca}
%\author{Nader Mobed}
%\email{nader.mobed@uregina.ca}
%\author{Giorgio Papini}
%\email{papini@uregina.ca}
%\altaffiliation[Also at ]{Prairie Particle Physics Institute, Regina, Saskatchewan, S4S 0A2, Canada;
%International Institute for Advanced Scientific Studies, 89019 Vietri sul Mare (SA), Italy.}
%\affiliation{Department of Physics, University of Regina \\ Regina, Saskatchewan, S4S 0A2, Canada}

\author{Dinesh Singh$^{a}$}
\altaffiliation[Electronic address:  ]{singhd@uregina.ca}
\author{Nader Mobed$^{a}$}
\altaffiliation[Electronic address:  ]{nader.mobed@uregina.ca}
\author{Giorgio Papini$^{a,b,c}$}
\altaffiliation[Electronic address:  ]{papini@uregina.ca}
\address{$^a$Department of Physics, University of Regina, Regina, Saskatchewan, S4S 0A2, Canada}
\address{$^b$Prairie Particle Physics Institute, Regina, Saskatchewan, S4S 0A2, Canada}
\address{$^c$International Institute for Advanced Scientific Studies, 89019 Vietri sul Mare (SA), Italy}

\date{\today}

\begin{abstract}

We show that spin-gravity interaction can distinguish between Dirac and Majorana
neutrino wave packets propagating in a Lense-Thirring background.
Using time-independent perturbation theory and gravitational phase to generate a perturbation Hamiltonian with spin-gravity coupling,
we show that the associated matrix element for the Majorana neutrino differs significantly from its Dirac counterpart.
This difference can be demonstrated through significant gravitational corrections to the neutrino oscillation length for
a two-flavour system, as shown explicitly for SN1987A.
%(Draft only.)

\end{abstract}

%\pacs{Valid PACS appear here}

\maketitle

%\author{D. Singh}
%\email{singhd@uregina.ca}
%\affiliation{%
%Department of Physics, University of Regina \\
%Regina, Saskatchewan, S4S 0A2, Canada
%}%
%\author{N. Mobed}
%\email{nader.mobed@uregina.ca}
%\affiliation{%
%Department of Physics, University of Regina \\
%Regina, Saskatchewan, S4S 0A2, Canada
%}%
%\author{G. Papini}
%\email{papini@uregina.ca}
%\affiliation{%
%Department of Physics, University of Regina \\
%Regina, Saskatchewan, S4S 0A2, Canada
%}%
%\affiliation{
%International Institute for Advanced Scientific Studies, 89019 Vietri sul Mare (SA), Italy.}
% %Lines break automatically or can be forced with \\


%\begin{abstract}

%\end{abstract}

%\maketitle

%\section{\label{sec:level1}First-level heading:\protect\\ The line
%break was forced \lowercase{via} \textbackslash\textbackslash}

%\subsection{\label{sec:level2}Second-level heading: Formatting}
%\subsubsection{\label{sec:level3}Third-level heading: References and Footnotes}

{\sl Introduction.--}An unresolved question in the Standard Model is whether neutrinos exist as Dirac or Majorana particles \cite{mohapatra,fukugita},
where the latter are regarded as the more natural candidates \cite{zee} to exist in Nature.
This is because the Majorana neutrino, as its own antiparticle, has only half the degrees of freedom compared
to the Dirac neutrino, and is deemed a more fundamental particle as a result.
%While a Dirac particle becomes a distinct antiparticle under charge conjugation,
%the Majorana particle is its own antiparticle.
%Majorana neutrinos also participate in neutrinoless double beta decay \cite{mohapatra,fukugita},
%while Dirac neutrinos do not.
One known process to distinguish between the two types is neutrinoless double beta
decay \cite{mohapatra,fukugita}, which only occurs for Majorana neutrinos.
However, direct observation of this phenomenon is at best inconclusive.
As well, Dirac and Majorana neutrinos are potentially distinguishable in magnetic fields,
since they possess unique magnetic moment structures \cite{mohapatra},
though this feature becomes relevant only when the fields are extremely strong.
%While Dirac and Majorana neutrinos have distinct magnetic moments \cite{mohapatra},
%their difference becomes relevant only for extremely strong magnetic fields.
A recent discovery \cite{Kam,SNO} has shown that neutrinos undergo flavour oscillations while propagating in vacuum,
inferring the existence of neutrino rest masses.
%This discovery provides the first concrete evidence for new physics beyond the Standard Model.
%However, the current theory of neutrino oscillations cannot distinguish between a
%Dirac and Majorana neutrino because only its left-handed chiral projection is subject to this
%phenomenon \cite{bilenky}, which happens to be identical for both types.
%However, the current theory cannot distinguish between Dirac and Majorana neutrinos because it only involves the
%left-handed chiral projection \cite{bilenky}, which is identical for both neutrino types.
However, the current theory of neutrino oscillations cannot distinguish between a Dirac and Majorana neutrino because only
its left-handed chiral projection is subject to this phenomenon \cite{bilenky}, which is identical for both types.

Although gravitational effects are usually neglected in particle physics, the fact that neutrinos are electrically neutral
provides an opportunity to study their long-range behaviour in response to curved space-time.
%It is reasonable to assume that certain quantum properties of neutrinos may be subject to gravitational effects.
For example, it is possible that a massive neutrino's helicity state can be flipped due to explicit coupling between its spin and
the background gravitational field.
If such an interaction exists, then we have means to probe the intrinsic nature of neutrinos due to gravity,
including the possibility to differentiate between Dirac and Majorana particles in a meaningful way.
By treating the two neutrino types as massive wave packets propagating in a Lense-Thirring (LT) background \cite{lense},
it is shown below that the Dirac and Majorana matrix elements
differ significantly, along with their respective oscillation lengths.

%As shown below, the Dirac and Majorana matrix elements and the corresponding oscillation lengths differ for the
%two neutrinos, which are treated as massive wave packets propagating in a Lense-Thirring (LT) background \cite{lense}.

%This is the underlying goal of this paper, which we consider for massive neutrino wave packets.
%propagating in a Lense-Thirring background \cite{lense}.
%, showing that the spin-gravity interaction leads to predicted corrections in the
%neutrino oscillation length that explicitly distinguish between Dirac and Majorana neutrinos.

%For all calculations presented here, we adopt geometric units
%$G = c = 1$ \cite{MTW}, such that all physical quantities are expressed in length units
%defined by $M$, the mass of the gravitational source, and the space-time metric has $-2$ signature.

{\sl Dirac Hamiltonian in Curved Space-Time.--}We begin with the four-dimensional covariant Dirac equation $\lt[i \gamma^\mu (x) D_\mu - {m \over \hbar}\rt]\psi (x) = 0$
of $-2$ signature $\lt(\mu = 0, 1, 2, 3\rt)$ with neutrino mass $m$, with $G = c = 1$ units~\cite{MTW}.
The curved space-time gamma matrices satisfying $\lt\{\gamma^\mu (x), \gamma^\nu (x)\rt\} = 2 \, g^{\mu \nu}(x)$
are expressed in terms of Minkowski gamma matrices $\gm^{\hat{\mu}}$ and
orthonormal vierbeins $\lt\{e^\mu{}_{\hat{\mu}}\rt\}$,
%onto a locally flat tangent space-time described by hatted indices
with $g_{\mu \nu} = \eta_{\hat{\alpha}\hat{\beta}} \, e^{\hat{\alpha}}{}_\mu \, e^{\hat{\beta}}{}_\nu, \quad
e^{\hat{\alpha}}{}_\mu \, e^\mu{}_{\hat{\beta}} = \delta^{\hat{\alpha}}{}_{\hat{\beta}}, \quad
e^{\mu}{}_{\hat{\alpha}} \, e^{\hat{\alpha}}{}_{\nu} \ = \ \delta^\mu{}_\nu$.
As well, $\gamma^\mu (x) = e^\mu{}_{\hat{\mu}} \gm^{\hat{\mu}}$ and $\lt\{\gamma^{\hat{\mu}}, \gamma^{\hat{\nu}} \rt\} = 2 \, \eta^{\hat{\mu} \hat{\nu}}$.
Then $D_\mu = \partial_\mu + i \, \Gm_\mu$ is the covariant derivative operator in terms of the spin connection
$\Gm_\mu = -{1 \over 4} \, \sigma^{\alpha \beta}(x) \, \Gamma_{\alpha \beta \mu} = -{1 \over 4} \, \sigma^{\hat{\alpha} \hat{\beta}} \,
\Gamma_{\hat{\alpha} \hat{\beta} \hat{\mu}} \, e^{\hat{\mu}}{}_\mu$,
where $\sigma^{\hat{\alpha} \hat{\beta}} = {i \over 2} [\gamma^{\hat{\alpha}}, \gamma^{\hat{\beta}}]$ and
$\Gamma_{\hat{\alpha} \hat{\beta} \hat{\mu}}$ are Ricci rotation coefficients.
The LT metric \cite{lense} for $x^\mu = \lt(t, x, y, z\rt)$ is
%
%\be
%\vec{g} & = &  \lt(1 - {2M \over r}\rt) \d t \otimes \d t
%\nn
%& &{} - \lt(1 + {2M \over r}\rt) \lt(\d x \otimes \d x + \d y \otimes \d y + \d z \otimes \d z \rt)
%\nn
%& &{} + {4 \over 5} \, {M \Omega R^2 \over r^3} \lt[x \lt(\d y \otimes \d t + \d t \otimes \d y \rt) \rt.
%\nn
%& &{} - \lt . y \lt(\d x \otimes \d t + \d t \otimes \d x \rt)\rt],
%\label{g-cart=}
%\ee
%
\be
\lefteqn{\vec{g} \ = \ \lt(1 - {2M \over r}\rt) \d t \otimes \d t
- \lt(1 + {2M \over r}\rt) \lt(\d x \otimes \d x  \rt. }
\nn
&&{} + \lt. \d y \otimes \d y + \d z \otimes \d z \rt)
+ {4 \over 5} \, {M \Omega R^2 \over r^3}
\nn
&&{} \times
\lt[x \lt(\d y \otimes \d t + \d t \otimes \d y \rt) - y \lt(\d x \otimes \d t + \d t \otimes \d x \rt)\rt], \qquad
\label{g-cart=}
\ee
with $M/r \ll 1$ and $M\Om R^2/r^2 \ll 1$,
where $r = \sqrt{x^2 + y^2 + z^2}$, $M$ and $R$ are the mass and radius of the gravitational source, and $\Om$ is its rotational frequency.
%Since (\ref{g-cart=}) describes a weak gravitational field,
The corresponding Dirac Hamiltonian to leading order in $M/r$ is
\be
\lefteqn{H_0 \ \approx \ \lt(1 - {2M \over r}\rt) \alvec \cdot \pvec
+ m \lt(1 - {M \over r}\rt) \bt + {i \hbar \, {M \over 2 \, r^3}} \lt(\alvec \cdot \rvec \rt) \qquad }
\nn
&&{}+ {4 \over 5} \, {M \Omega R^2 \over r^3} \, \Lvec^{\Cz} + {1 \over 5} \, {\hbar \, M \Omega R^2 \over r^3}
\lt[{3 \, z \over r^2} \lt(\Sgvec \cdot \rvec\rt) - \Sgvec^{\Cz}\rt], \qquad
%\nn
\label{H0=}
\ee
where $\alvec$ and $\bt$ are the Dirac matrices, $\Sgvec^{\Cj} = \sg^{\ze \Cj}$ is the $x^j$-component of the spin angular momentum operator,
and $\Lvec^{\Cz}$ is the orbital angular momentum operator in the $z$-direction.
In spherical co-ordinates, the field point is $\rvec = \lt(r, \th, \vph\rt)$, defined in relation to a cartesian co-ordinate frame
expressed by $x^\mu$.
The energy eigenvalue for
$H_0 \lt| \psi_0 \rt\rangle = E_0^{(\pm)} \lt| \psi_0 \rt\rangle$ is
\be
E_0^{(\pm)} & \approx & \sqrt{(\hbar \, \k)^2 + m^2} - {2M \over r} \, (\hbar \, \k)
\nn
& &{}
+ {4 \over 5} \,  {M \Omega R^2\over r^3} \lt(L^{\hat{z}} \pm {\hbar \over 2}\rt), \quad
\label{E0=}
\ee
where $p = \hbar \, \k$ is the neutrino's momentum eigenvalue.

%By themselves, (\ref{H0=}) and (\ref{E0=}) do not provide any new insights involving gravitation because the only term in $E_0^{(\pm)}$
%which contains the neutrino mass is purely the special relativistic energy, which is decoupled from the gravitational field.
%However, there is another way in which gravitation can enter non-trivially into this quantum process, and that is through
%the introduction of a {\it gravitational phase} in the neutrino's wavefunction.
%Previous work has shown \cite{cai,singh,papini1,papini2} that classical gravitational fields can be incorporated into the dynamics of a quantum particle by means of
%a gauge invariant phase function $\PhiG$, where for the metric deviation $h_{\mu \nu} = g_{\mu \nu} - \eta_{\mu \nu} \ll 1$,
%

{\sl Gravitational Phase.--}By itself, (\ref{H0=}) is insufficient to describe a spin-1/2 particle interaction with gravity.
However, for a weak field described by $h_{\mu \nu} = g_{\mu \nu} - \eta_{\mu \nu} \ll 1$,
a gauge invariant gravitational phase
\be
\PhiG & \equiv & {1 \over 2} \int_{x_0^\mu}^{x^\mu} \d z^\lm
h_{\lm \al} (z) \pvec^\al
\nn
& &{}
- {1 \over 4} \int_{x_0^\mu}^{x^\mu} \d z^\lm \lt[h_{\lm \al, \bt} (z) - h_{\lm \bt, \al} (z) \rt]
\vec{L}^{\alpha \beta} (z) \quad
\nn
& = & \int_{t_0}^{t} \, \d t' \, \lt(\nabvec_t \PhiG\rt) +
\int_{x_0}^{x} \, \d x' \, \lt(\nabvec_x \PhiG\rt)
\nn
& &{} +
\int_{y_0}^{y} \, \d y' \, \lt(\nabvec_y \PhiG\rt) +
\int_{z_0}^{z} \, \d z' \, \lt(\nabvec_z \PhiG\rt)
%
%& = & \int_{\rvec_0}^{\rvec} \, \d z^\mu \, \lt(\nabvec_\mu \PhiG\rt)
\label{PhiG=}
\ee
defined along some space-time trajectory $z^\mu = \lt(t', x', y', z'\rt)$ leads to a weak-field solution of the covariant
Klein-Gordon equation \cite{cai,singh,papini1,papini2}, where $\pvec^\al$ and $\Lvec^{\al \bt}$ are the generators of linear and orbital angular momentum
for a free particle.
Use of (\ref{PhiG=}) according to the phase transformation $\psi(x) \rightarrow \exp\lt(i \PhiG/\hbar\rt) \psi(x)$ results in a
new Dirac Hamiltonian $H = H_0 + H_{\PhiG}$, where
\be
H_{\PhiG} & = & \alvec \cdot \lt(\nabvec \PhiG \rt) + \lt(\nabvec_t \PhiG \rt)
\label{H-phiG=}
\ee
is treated as a perturbation of $H_0$,
reproducing all terms that have been observed, derived, or predicted to exist \cite{hehl}
for a spin-1/2 particle in a gravitational field.
%It follows that (\ref{H-phiG=}) and a wave packet treatment of neutrinos can lead to a perturbed energy that is dependent on mass terms coupled to the
%gravitational field.
%While a previous investigation following a semianalytic approach \cite{singh1} is available for review, this paper
%offers a much more complete analysis with almost no approximations employed.

{\sl Neutrino Wave Packets.--}To formulate the wave packet description, we begin with a wavefunction composed of a linear superposition of
plane waves \cite{singh1}
\be
| \psi \rangle & = &
{1  \over (2\pi)^{3/2}} \int \d^3 k \, \xi(\vec{k}) \,
e^{i \vec{k} \cdot \vec{r}} |U(\vec{k})\rangle,
\label{psi=}
\ee
where $e^{i \vec{k} \cdot \vec{r}} |U(\vec{k})\rangle$ is a normalized solution of the free-particle Dirac equation, and
\be
\xi(\kvec) & = & {1 \over ({\sqrt{2\pi} \, \sigma_{\rm p}})^{3/2}} \, \exp\lt[- {(\kvec - \kvec_0)^2 \over 4 \, \sigma_{\rm p}^2} \rt]
\label{Gaussian=}
\ee
is a Gaussian function in momentum space, of width $\sigma_{\rm p}$ and centroid $\kvec_0$, with $\k = |\kvec_0|$.
The matrix element due to (\ref{H-phiG=}) is then
\be
\lefteqn{\langle \psi(\rvec) |H_{\PhiG}| \psi(\rvec) \rangle \ = \
{1 \over (2\pi)^3} \int \d^3 r' \ \d^3 k \ \d^3 k' \ \xi(\kvec) \, \xi(\kvec')  }
\nn
&&{} \times  \exp\lt[i\lt(\kvec - \kvec'\rt)\cdot \rvec'\rt] \langle U(\kvec') | H_{\PhiG}(\rvec, \rvec') | U(\kvec) \rangle, \qquad
\label{matrix-element}
\ee
where the integration is performed over all phase space in spherical co-ordinates,
excluding the region occupied by the gravitational source.
To evaluate (\ref{matrix-element}) explicitly, we need to specify $| U(\kvec) \rangle$.
Assuming the Weyl representation \cite{fukugita} for the gamma matrices, it is understood that
the Dirac four-spinor is
\be
| U(\kvec) \rangle^{\rm Dirac} & = & | \nu_L \rangle + | \nu_R \rangle,
\label{U-Dirac}
\ee
where $| \nu_{L(R)} \rangle$ is its left- (right)-handed chiral projection.
In contrast, a Majorana four-spinor is identical to itself (up to some phase) under charge conjugation \cite{fukugita},
where $| \chi^c \rangle = \pm | \chi \rangle$.
Then
\be
| U(\kvec) \rangle^{\rm Maj.} & = & \lt\{
\begin{array}{ccc}
| W_1(\kvec) \rangle^{\rm Maj.} & \equiv & | \nu_L \rangle + | \nu_L^c \rangle, \\ \\
| W_2(\kvec) \rangle^{\rm Maj.} & \equiv & | \nu_R \rangle - | \nu_R^c \rangle,
\end{array} \rt.
\label{pm-mass-eigenstates}
\ee
and $| W_{1(2)}^c \rangle^{\rm Maj.} = \pm | W_{1(2)} \rangle^{\rm Maj.}$.

After substituting (\ref{U-Dirac}) or (\ref{pm-mass-eigenstates}) into (\ref{matrix-element}), we proceed to
evaluate the Dirac or Majorana matrix element with the Rayleigh plane wave expansion \cite{arfken}
%
%\be
%e^{i \, \kvec \cdot \rvec} & = & 4 \, \pi \sum_{l=0}^\infty \, \sum_{m=-l}^l \, i^l \, j_l\lt(k \, r\rt) Y_{lm}^*\lt(\lm,\mu\rt) \, Y_{lm}\lt(\th,\vph\rt),
%\qquad
%\label{Rayleigh}
%\ee
%
in terms of spherical Bessel functions and spherical harmonics defined for both position and momentum space angles.
It follows that the orthonormality conditions serve to truncate the series expansion, allowing a virtually {\em exact}
evaluation of the matrix element.
Generically, (\ref{matrix-element}) is the sum of both spin-diagonal terms and spin-flip terms
proportional to the Pauli spin matrices $\sgvec^{\Cj}$.
Since the neutrino's spin quantization axis is parallel to the direction of propagation,
the helicity transition element is \cite{singh1}
\be
\langle \pm | \sgvec | \mp \rangle & = &
\lt[\cos \th \, \cos \vph \pm i \, \sin \vph \rt] \vec{\hat{x}}
\nn
& &{} + \lt[\cos \th \, \sin \vph \mp i \, \cos \vph \rt] \vec{\hat{y}} -
\sin \th \, \vec{\hat{z}}, \qquad
\label{orientation=}
\ee
where $| \pm \rangle$ are the two-component spinors which define positive (negative) helicity
for the neutrino.

\begin{figure}
\psfrag{q}[cc][][2.5][0]{\hspace{0.5cm} $q$ (dimensionless)}
\psfrag{C0}[bc][][2.5][0]{$C_0$ (dimensionless)}
\psfrag{C1}[bc][][2.5][0]{$C_1$ (dimensionless)}
\psfrag{C2}[bc][][2.5][0]{$C_2$ (dimensionless)}
\psfrag{C0y}[bc][][2.5][0]{$C_{0\hat{y}}$ (dimensionless)}
\psfrag{C1y}[bc][][2.5][0]{$C_{1\hat{y}}$ (dimensionless)}
\psfrag{C2y}[bc][][2.5][0]{$C_{2\hat{y}}$ (dimensionless)}
\psfrag{D0}[bc][][2.5][0]{$D_0$ (dimensionless)}
\psfrag{D1}[bc][][2.5][0]{$D_1$ (dimensionless)}
\psfrag{D2}[bc][][2.5][0]{$D_2$ (dimensionless)}
\psfrag{D0x}[bc][][2.5][0]{$D_{0\hat{x}}$ (dimensionless)}
\psfrag{D1x}[bc][][2.5][0]{$D_{1\hat{x}}$ (dimensionless)}
\psfrag{D2x}[bc][][2.5][0]{$D_{2\hat{x}}$ (dimensionless)}
\psfrag{F1_Dirac}[bc][][2.5][0]{$F_1^{\rm Dirac}$ (dimensionless)}
\psfrag{F2_Dirac}[bc][][2.5][0]{$F_2^{\rm Dirac}$ (dimensionless)}
\psfrag{F1_Maj}[bc][][2.5][0]{$F_1^{\rm Maj.}$ (dimensionless)}
\psfrag{F2_Maj}[bc][][2.5][0]{$F_2^{\rm Maj.}$ (dimensionless)}
\psfrag{th = 0.5*Pi}[cc][][2.5][0]{\small $\th = 0.5 \pi$}
\psfrag{th = 0.4*Pi}[cc][][2.5][0]{\small $\th = 0.4 \pi$}
\psfrag{th = 0.3*Pi}[cc][][2.5][0]{\small $\th = 0.3 \pi$}
\psfrag{th = 0.2*Pi}[cc][][2.5][0]{\small $\th = 0.2 \pi$}
\psfrag{th = 0.1*Pi}[cc][][2.5][0]{\small $\th = 0.1 \pi$}
\psfrag{th = 0}[cc][][2.5][0]{\small $\th = 0$}
\begin{minipage}[t]{0.3 \textwidth}
\centering
\subfigure[\hspace{0.2cm} $F_1^{\rm Dirac}$ \, (SN1987A)]{
\label{fig:F1-SN1987A-Dirac}
%\rotatebox{0}{\includegraphics[width = 6cm, height = 4.5cm, scale = 1]{F1-SN1987A-Dirac}}}
\rotatebox{0}{\includegraphics[width = 6cm, height = 4.5cm, scale = 1]{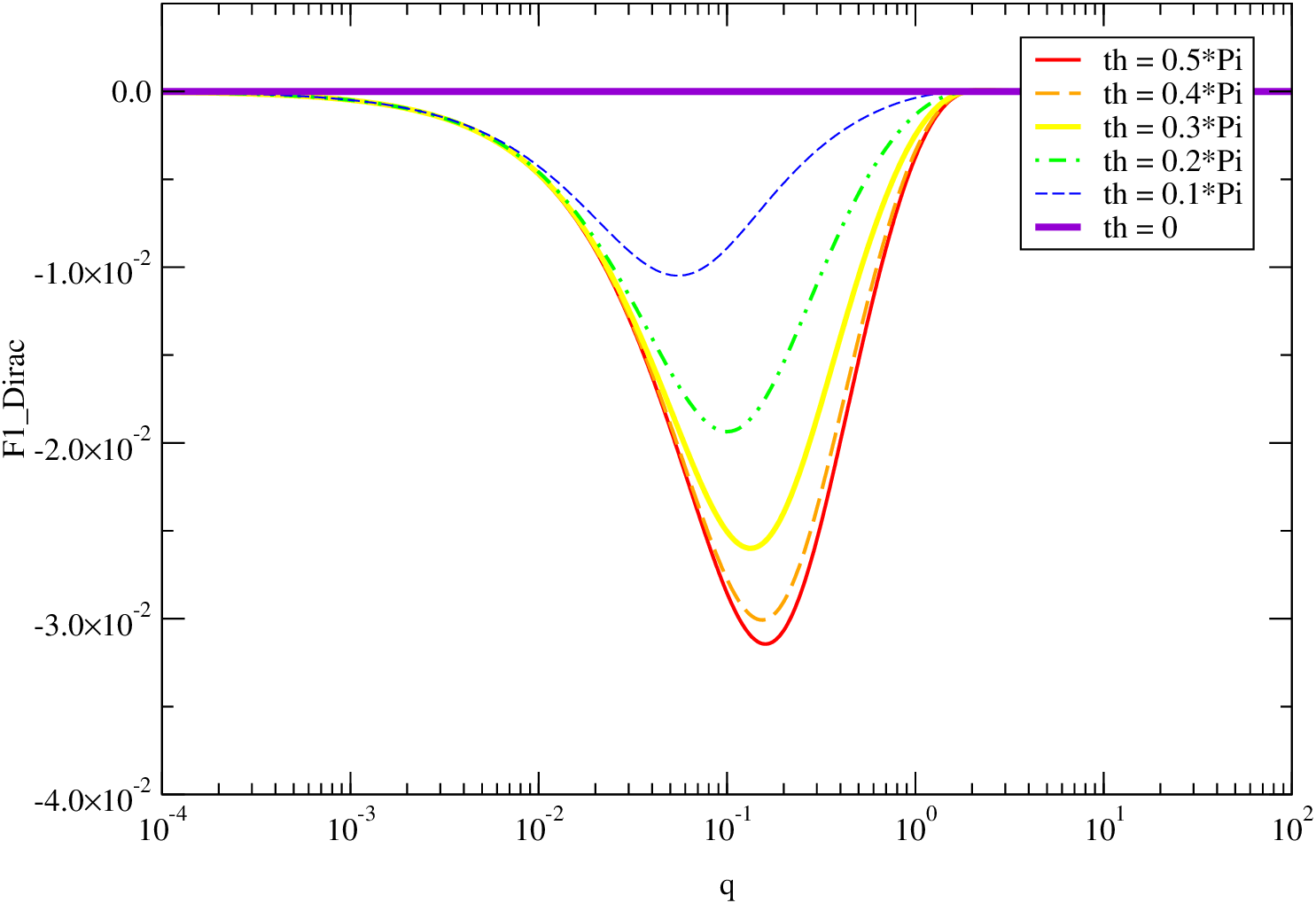}}}
\vspace{0.5cm}
\end{minipage}%
\hspace{2.0cm}
\begin{minipage}[t]{0.3 \textwidth}
\centering
\subfigure[\hspace{0.2cm} $F_1^{\rm Maj.}$ \, (SN1987A)]{
\label{fig:F1-SN1987A-Maj}
%\rotatebox{0}{\includegraphics[width = 6cm, height = 4.5cm, scale = 1]{F1-SN1987A-Maj}}}
\rotatebox{0}{\includegraphics[width = 6cm, height = 4.5cm, scale = 1]{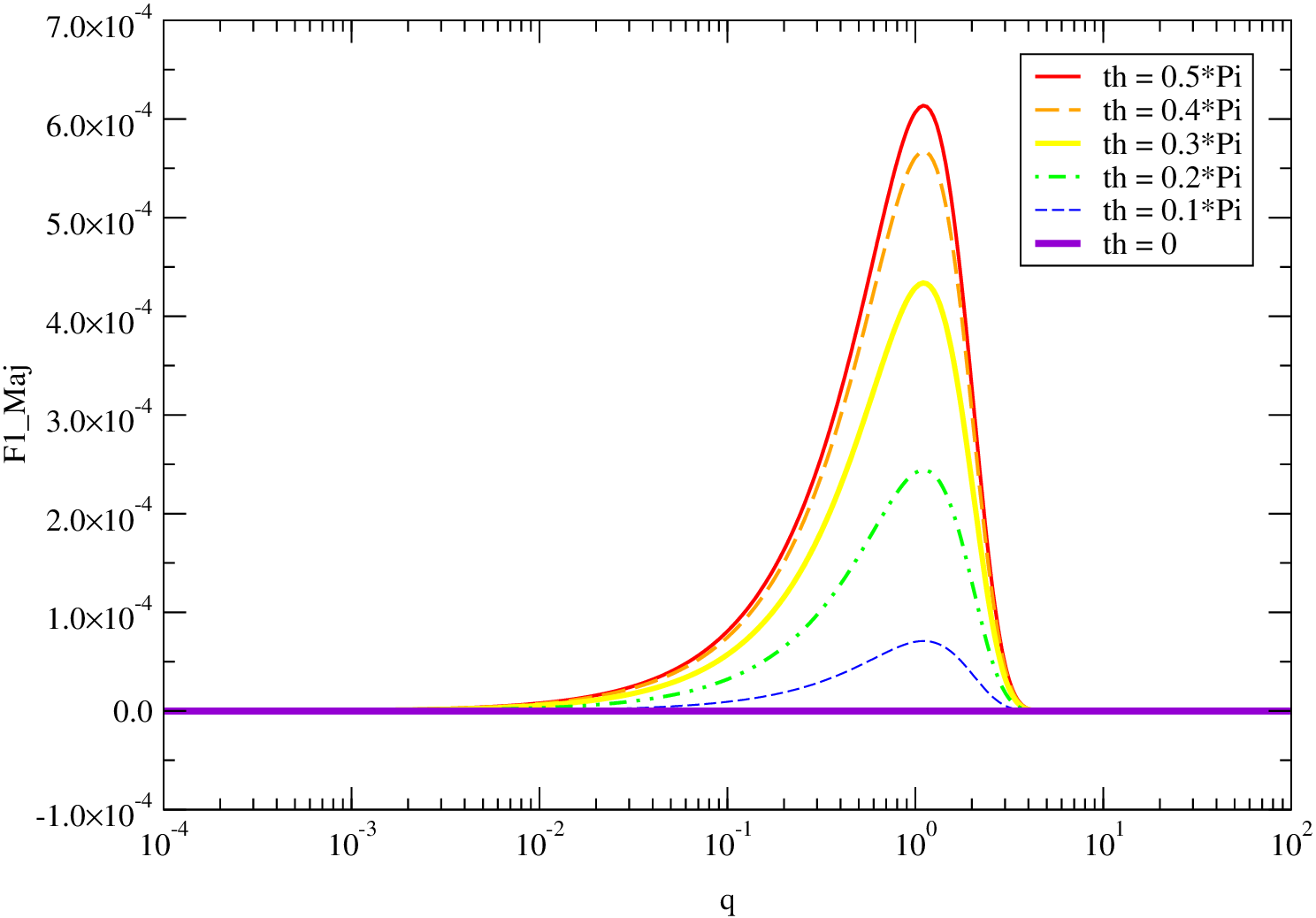}}}
\end{minipage}
\caption{\label{fig:F1-SN1987A}  $F_1$ as a function of $q$ due to the SN1987A gravitational source,
for varying neutrino beam angle $\th$.
Besides having opposite sign, $F_1^{\rm Dirac}$ is two orders of magnitude larger than $F_1^{\rm Maj.}$.}
\end{figure}
{\sl Dirac and Majorana Matrix Elements.--}After performing a power series expansion of (\ref{matrix-element}) with respect to $\bar{m} \equiv m/(\hbar \, \k) \ll 1$,
we present the main formal results.
For the Dirac neutrino, the gravitational phase-induced matrix element is
\be
\lefteqn{\langle \psi(\vec{r}) |H_{\Phi_{\rm G}}| \psi(\vec{r}) \rangle^{\rm Dirac} \ = \ }
\nn
&&{} (\hbar \, \k) \lt\{
{M \over r} \lt[C_0 + C_1 \, \bar{m} + C_2 \, \bar{m}^2 \rt] \rt.
\nn
&&{} + \lt. {M \Omega R^2\over r^2} \sin \theta \lt[D_0 + D_1 \, \bar{m} + D_2 \, \bar{m}^2 \rt] \rt\},
\label{amplitude-Dirac}
\ee
where $C_j$ and $D_j$ are dimensionless functions of $\k$, $R$, $r$, and $q \equiv \k/\sg_{\rm p}$,
whose explicit expressions are shown in a much longer paper \cite{singh2}.
Some important details in (\ref{amplitude-Dirac}) are as follows.
First, $C_j$ correspond to the spin-diagonal parts of (\ref{matrix-element})
coupled to $M/r$, while $D_j$ refer to the spin-flip parts coupled to $M \Om R^2/r^2$.
Second, the presence of $\sin \th$ clearly indicates that only terms with the $z$-component of (\ref{orientation=})
survive the integration.
The most obvious interpretation is that the gravitational source's rotation induces the
helicity transition of the Dirac neutrino for propagation away from the axis of symmetry.
That is, the off-diagonal metric terms in (\ref{g-cart=}) resemble an inhomogeneous magnetic field generating
the spin-flip of a particle.
Third, there are terms {\em linear} in $\bar{m}$ present in (\ref{amplitude-Dirac}) from
the fact that the normalization coefficient in $| U(\kvec) \rangle^{\rm Dirac}$ is $\sqrt{(E + m)/(2E)}$, where $E = \sqrt{p^2 + m^2}$,
with important consequences to follow.

Evaluating (\ref{matrix-element}) for the Majorana neutrino, we have
\be
\lefteqn{\langle \psi_{1(2)}(\vec{r}) |H_{\Phi_{\rm G}}| \psi_{1(2)}(\vec{r}) \rangle^{\rm Maj.} \ = \
%\langle \psi(\vec{r}) |H_{\Phi_{\rm G}}| \psi(\vec{r}) \rangle^{\rm Dirac}
}
\nn
%& &{}  - (\hbar \, \k) \, {M \Omega R^2\over r^2} \sin \theta \lt[D_0 + D_1 \, \bar{m} + D_2 \, \bar{m}^2 \rt]
& &{} (\hbar \, \k) \lt\{ {M \over r} \lt[C_0 + C_1 \, \bar{m} + C_2 \, \bar{m}^2 \rt] \rt.
\nn
&&{} \pm \sin \th \, \sin \vph
%\nn
%&&{} \times
\lt[ {M \over r}
\langle \pm |\sgvec|\mp \rangle^{\Cy} \lt[C_{0 \Cy} + C_{1 \Cy} \, \bar{m} + C_{2 \Cy} \, \bar{m}^2\rt] \rt.
\nn
&&{} + \lt. \lt. {M \Omega R^2\over r^2} \langle \pm |\sgvec|\mp \rangle^{\Cx} \lt[D_{0 \Cx} + D_{1 \Cx} \, \bar{m} + D_{2 \Cx} \, \bar{m}^2\rt] \rt] \rt\},
\qquad
%\nn
\label{amplitude-Maj}
\ee
where the ``1(2)'' refers to the upper (lower) signs in (\ref{amplitude-Maj}), and
$C_{j\Cy}$ and $D_{j\Cx}$ are also dimensionless functions of $\k$, $R$, $r$, and $q$.
The differences between (\ref{amplitude-Dirac}) and (\ref{amplitude-Maj}) are quite striking.
First, while both have the spin-diagonal terms $C_j$,
the spin-flip terms have an overall factor of $\sin \th \, \sin \vph$.
This corresponds to the $y$-component of the neutrino beam, a direct consequence of the Majorana neutrino's self-conjugation condition,
as the charge conjugation operation involves the presence of $\sgvec^{\Cy}$ in its definition \cite{fukugita}.
This result suggests a preferred direction orthogonal to the source's axis of symmetry.
However, since the LT metric is {\em axisymmetric}, this $\vph$-dependence on (\ref{amplitude-Maj})
purely results from how we defined the co-ordinate system beforehand.
This anisotropy can be removed by averaging over a complete cycle.
Even so, if the source has a significant quadrupole moment to induce an azimuthal perturbation
of the LT metric, then a resonance effect may be generated under suitable conditions.
Second, the spin-flip parts of (\ref{amplitude-Maj}) are dependent on the $x$- and $y$-components of
(\ref{orientation=}), as opposed to the $z$-component for the Dirac neutrino.
Third, a spin-flip term still contributes to the Majorana matrix element in the limit as $\Om \rightarrow 0$,
while no such term survives for the Dirac counterpart.
Though this seems counterintuitive, terms of this type are expected to be present
because of the self-conjugate nature of Majorana neutrinos.
%A detailed discussion of this observation is found in the upcoming longer paper \cite{singh2}.

%This difference is quite counterintuitive, so it is unclear why this extra helicity transition effect
%should appear here.

\begin{figure}
\psfrag{q}[cc][][2.5][0]{\hspace{0.5cm} $q$ (dimensionless)}
\psfrag{C0}[bc][][2.5][0]{$C_0$ (dimensionless)}
\psfrag{C1}[bc][][2.5][0]{$C_1$ (dimensionless)}
\psfrag{C2}[bc][][2.5][0]{$C_2$ (dimensionless)}
\psfrag{C0y}[bc][][2.5][0]{$C_{0\hat{y}}$ (dimensionless)}
\psfrag{C1y}[bc][][2.5][0]{$C_{1\hat{y}}$ (dimensionless)}
\psfrag{C2y}[bc][][2.5][0]{$C_{2\hat{y}}$ (dimensionless)}
\psfrag{D0}[bc][][2.5][0]{$D_0$ (dimensionless)}
\psfrag{D1}[bc][][2.5][0]{$D_1$ (dimensionless)}
\psfrag{D2}[bc][][2.5][0]{$D_2$ (dimensionless)}
\psfrag{D0x}[bc][][2.5][0]{$D_{0\hat{x}}$ (dimensionless)}
\psfrag{D1x}[bc][][2.5][0]{$D_{1\hat{x}}$ (dimensionless)}
\psfrag{D2x}[bc][][2.5][0]{$D_{2\hat{x}}$ (dimensionless)}
\psfrag{F1_Dirac}[bc][][2.5][0]{$F_1^{\rm Dirac}$ (dimensionless)}
\psfrag{F2_Dirac}[bc][][2.5][0]{$F_2^{\rm Dirac}$ (dimensionless)}
\psfrag{F1_Maj}[bc][][2.5][0]{$F_1^{\rm Maj.}$ (dimensionless)}
\psfrag{F2_Maj}[bc][][2.5][0]{$F_2^{\rm Maj.}$ (dimensionless)}
\psfrag{th = 0.5*Pi}[cc][][2.5][0]{\small $\th = 0.5 \pi$}
\psfrag{th = 0.4*Pi}[cc][][2.5][0]{\small $\th = 0.4 \pi$}
\psfrag{th = 0.3*Pi}[cc][][2.5][0]{\small $\th = 0.3 \pi$}
\psfrag{th = 0.2*Pi}[cc][][2.5][0]{\small $\th = 0.2 \pi$}
\psfrag{th = 0.1*Pi}[cc][][2.5][0]{\small $\th = 0.1 \pi$}
\psfrag{th = 0}[cc][][2.5][0]{\small $\th = 0$}
\begin{minipage}[t]{0.3 \textwidth}
\centering
\subfigure[\hspace{0.2cm} $F_2^{\rm Dirac}$ \, (SN1987A)]{
\label{fig:F2-SN1987A-Dirac}
%\rotatebox{0}{\includegraphics[width = 6cm, height = 4.5cm, scale = 1]{F2-SN1987A-Dirac}}}
\rotatebox{0}{\includegraphics[width = 6cm, height = 4.5cm, scale = 1]{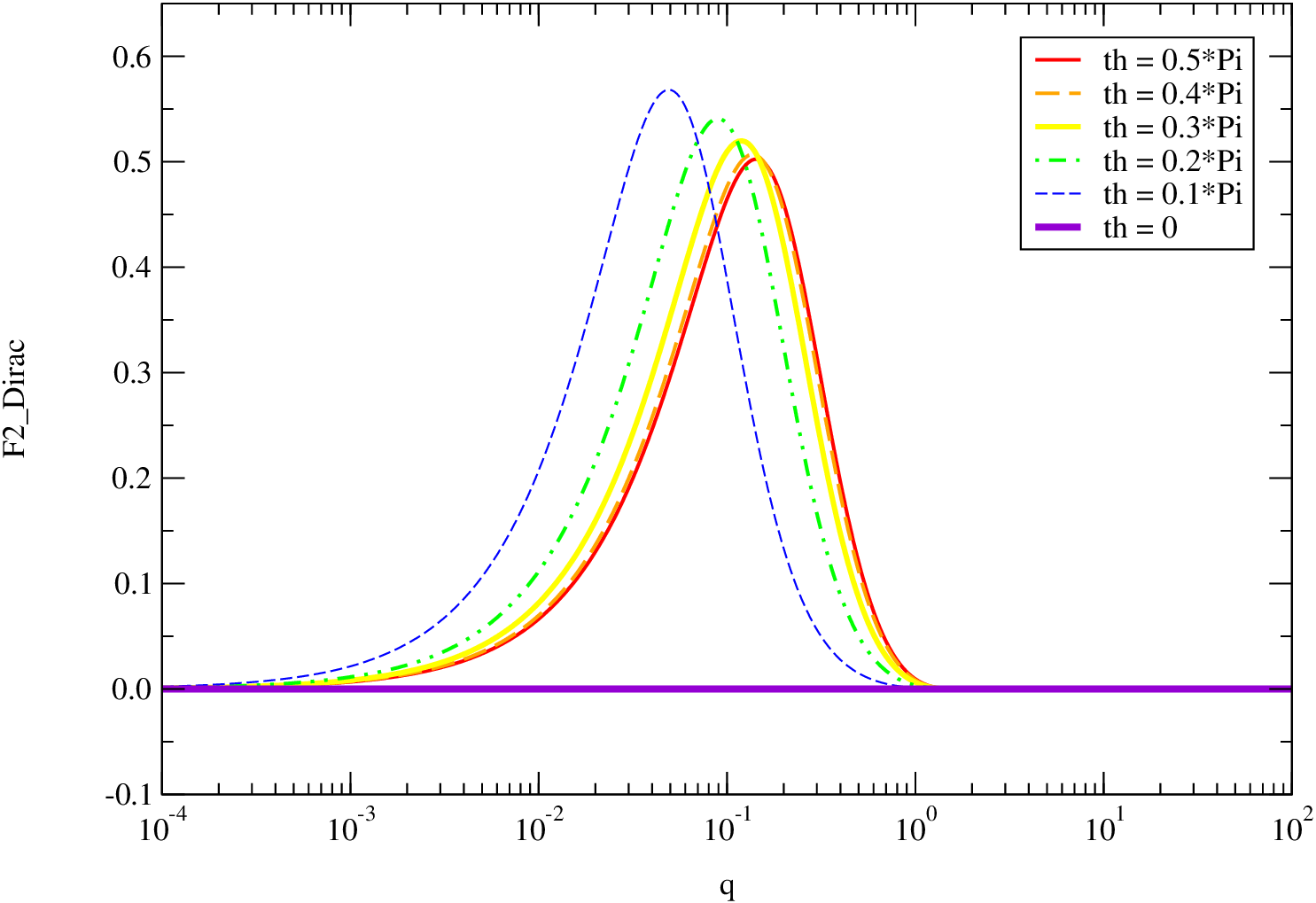}}}
\vspace{0.5cm}
\end{minipage}%
\hspace{2.0cm}
\begin{minipage}[t]{0.3 \textwidth}
\centering
\subfigure[\hspace{0.2cm} $F_2^{\rm Maj.}$ \, (SN1987A)]{
\label{fig:F2-SN1987A-Maj}
%\rotatebox{0}{\includegraphics[width = 6cm, height = 4.5cm, scale = 1]{F2-SN1987A-Maj}}}
\rotatebox{0}{\includegraphics[width = 6cm, height = 4.5cm, scale = 1]{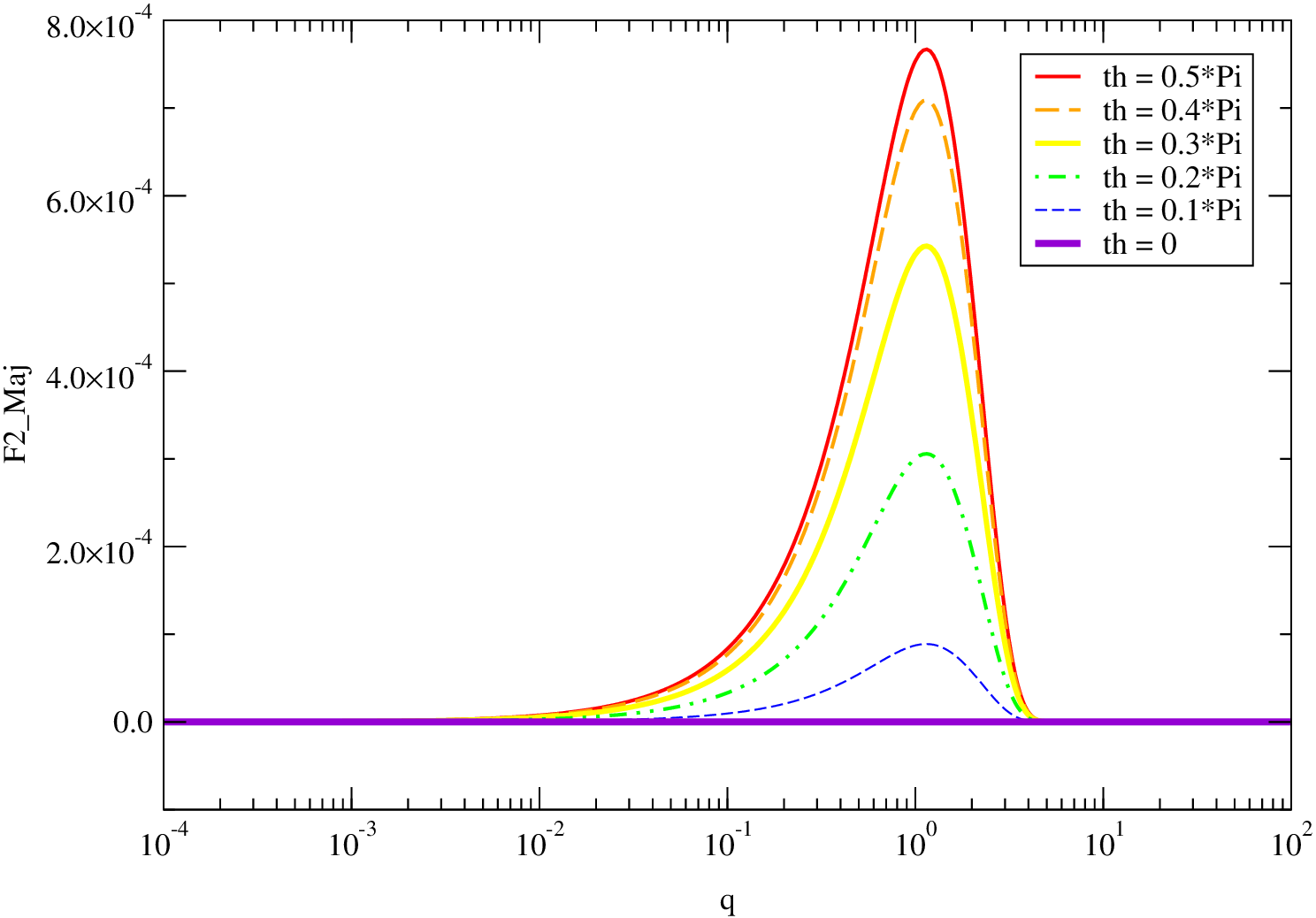}}}
\end{minipage}
\caption{\label{fig:F2-SN1987A}  $F_2$ as a function of $q$ applied to SN1987A for varying $\th$.
The gravitational correction for $F_2^{\rm Dirac}$ is larger than $F_2^{\rm Maj.}$ by three orders of magnitude.}
\end{figure}

{\sl Spin-Gravity Corrections to Neutrino Oscillation Length.--}We now demonstrate how the matrix elements (\ref{amplitude-Dirac}) and (\ref{amplitude-Maj}) lead
to predicted gravitational corrections in the neutrino oscillation length, defined as
$L_{\rm osc.} = 2 \pi/\lt(E_{\bar{m}_2}^{(\pm)} - E_{\bar{m}_1}^{(\pm)}\rt)$ \cite{mohapatra,fukugita}, where by convention we set
$m_2 > m_1$.
To obtain $E_{\bar{m}_j}^{(\pm)}$, we use the Brillouin-Wigner (BW) method \cite{ballentine},
instead of the more familiar Rayleigh-Schr\"{o}dinger (RS) method \cite{sakurai}.
From the BW approach applied to a second-order perturbation, we have
\be
E_{\bar{m}_j}^{(\pm)} & = & E_0^{(\pm)} + \lt\langle \pm \rt| H_{\PhiG} \lt| \pm \rt\rangle
+ {\lt|\lt\langle \mp \rt| H_{\PhiG} \lt| \pm \rt\rangle\rt|^2 \over E_{\bar{m}_j}^{(\pm)} - E_0^{(\mp)}},
\label{E-perturbed1=}
\ee
where the unperturbed energy $E_0^{(\pm)}$ is described by (\ref{E0=}), and
$\lt\langle \pm \rt| H_{\PhiG} \lt| \pm \rt\rangle$ and $\lt\langle \mp \rt| H_{\PhiG} \lt| \pm \rt\rangle$
are the spin-diagonal and spin-flip components of the Dirac and Majorana matrix elements, as found in
(\ref{amplitude-Dirac}) and (\ref{amplitude-Maj}), respectively.

%
%\be
%\lt| \pm \rt\rangle & \equiv & \lt\{
%\begin{array}{cl}
%| U(\kvec) \rangle^{\rm Dirac}_{\pm} & \qquad {\rm (Dirac),} \\ \\
%| W_1(\kvec) \rangle^{\rm Maj.}_{\pm} & \qquad {\rm (Majorana),}
%\end{array} \rt.
%\label{pm-mass-eigenstates}
%\ee
%

%Although the presence of $E_{\bar{m}_j}^{(\pm)}$ in the denominator of (\ref{E-perturbed1=}) makes the BW approach
%generally impossible to solve explicitly for the total energy,
The advantage of the BW method comes from knowing that it
yields an {\em exact} expression for the total energy involving a two-level spin system,
which is precisely what we have.
Therefore, we can solve for $E_{\bar{m}_2}^{(\pm)} - E_{\bar{m}_1}^{(\pm)}$
after averaging over the azimuthal angular dependence in the helicity transition term via
\be
\lt| \lt\langle \mp \rt| H_{\PhiG} \lt| \pm \rt\rangle \rt|^2 & \rightarrow & {1 \over 2 \pi} \int_0^{2\pi} \lt| \lt\langle \mp \rt| H_{\PhiG} \lt| \pm \rt\rangle \rt|^2 \, \d \vph.
\label{spinflip-avg-Maj}
\ee
This leads to the expression
\be
\lefteqn{E_{\bar{m}_2}^{(\pm)} - E_{\bar{m}_1}^{(\pm)} \ = \ }
\nn
&& (\hbar \, \k) \lt[F_1 \lt(\bar{m}_2 -  \bar{m}_1\rt) + \lt(F_2 + {1 \over 2}\rt) \lt(\bar{m}_2^2 -  \bar{m}_1^2\rt) \rt], \qquad
\label{Energy-shift=}
\ee
where $F_1$ and $F_2$ are also dimensionless functions of $\k$, $R$, $r$, and $q$,
and dependent on a complicated combination of terms coupled to $M/r$ and $M\Om R^2/r^2$,
whose analytic expressions are deferred to the forthcoming longer paper \cite{singh2}.
As a test case, we present $F_1$ and $F_2$ as a function of $q$ in Figures~\ref{fig:F1-SN1987A} and \ref{fig:F2-SN1987A}, respectively,
for varying orientations of the neutrino beam angle $\th$, using data from SN1987A \cite{soida}, where $M \approx 1.4 \, M_\odot$,
$R \approx 10$ km, $\Om \approx 2.936$ kHz, and $r \approx 49$ kpc.
The neutrino wave packet is assumed to have a mean momentum of $\hbar \, \k = 1$ MeV throughout.

For all plots of $F_j$ considered, the differences between positive and negative helicity are negligibly small.
Also, the functions are insensitive to moderate or large changes in $\k$.
%A thorough analysis of $F_j$ for both SN1987A and the Sun is found in the longer paper
%to follow ~\cite{singh2}, but we can make some general remarks about the plots here.
Comparing Figures~\ref{fig:F1-SN1987A-Dirac} and \ref{fig:F1-SN1987A-Maj} for $\th = \pi/2$, we see that
$F_1^{\rm Dirac}$ and $F_1^{\rm Maj.}$ have opposite sign, and $|F_1^{\rm Dirac}|$ peaks near
$3 \times 10^{-2}$, while the corresponding maximum $F_1^{\rm Maj.}$ is roughly $6 \times 10^{-4}$.
Regarding $F_2$, Figures~\ref{fig:F2-SN1987A-Dirac} and \ref{fig:F2-SN1987A-Maj} for $\th = \pi/2$ show that
$F_2^{\rm Dirac}$ has a maximum value near $0.5$, while $F_2^{\rm Maj.}$ peaks near $8 \times 10^{-4}$,
about three orders of magnitude smaller.
The fact that $F_j$ are non-zero for the range of $10^{-4} \lesssim q \lesssim 10^1$ is
consistent with reasonable choices of $\sg_{\rm p}$ for neutrinos produced in a neutron star, based on a
mean free path calculation assuming known stellar data \cite{singh1}.

{\sl Conclusion.--}This paper indicates that Dirac and Majorana wave packets interact differently with
gravity, with potential observational consequences.
It is especially valuable to realize that the SN1987A values for $F_1$, while small, are {\em not} negligible.
This presents the interesting possibility that we can extract observational knowledge of the absolute mass difference $m_2 - m_1$
for a two-flavour oscillating system.
Use of (\ref{Energy-shift=}) can then lead to a determination of the absolute neutrino masses $m_1$ and $m_2$
by a parameter fit of $q$, $\bar{m}_2 -  \bar{m}_1$, and $\bar{m}_2^2 -  \bar{m}_1^2$ to precision measurements of the
neutrino oscillation length, should this possibility become accessible in the future.

%It seems that further study of this possibility merits attention.
%{\sl Conclusion.--}This paper suggests that Dirac and Majorana neutrinos are distinguishable via spin-gravity
%interaction, with potential observational consequences worth investigating in detail.

{\sl Acknowledgements.--}This research is supported, in part, by the Natural Sciences and Engineering Research Council of Canada (NSERC).

%%%%%%%%%%%%%%%%%%%%%%%%%%%%%%%%%%%%%%%%%%%%%%%%%%%%%%%%%%%%%%%%%%%%%%%%%%%%%%%%%%%%%%%%%%%%%%%

\end{document}